# Observation of Arbitrarily Configurable Nonlinear Topological Modes

Kai Bai,[1] Chen Lin,[1] Jia-Zheng Li,[1] Tao Liu,[1] Duanduan Wan,[1*] Meng Xiao,[1, 2*]

[1]Key Laboratory of Artificial Micro- and Nano-structures of Ministry of Education and School of Physics and Technology, Wuhan University, Wuhan 430072, China

[2]Wuhan Institute of Quantum Technology, Wuhan 430206, China

Corresponding Email [*]: phmxiao@whu.edu.cn; ddwan@whu.edu.cn;

**Abstract:**

Nonlinear topology is an emerging field that combines the intrinsic reconfigurability of nonlinear systems with the robustness of topological protection, offering fertile ground for unconventional phenomena and novel applications. Recently, arbitrarily configurable nonlinear topological modes (ANTMs) were proposed, enabling wavefunctions to be configured into arbitrary profiles , and offering greatly enhanced capacity for topological modes and high-throughput topological transport. Here we present the first direct experimental demonstration of ANTMs. These nonlinear topological modes are robust against disorder while also being continuously reshaped and reconfigured in real time through external control. These counterintuitive properties highlight the versatility of arbitrarily morphing nonlinear topological modes and pave the way for highly adaptable topological devices capable of operating reliably across diverse application scenarios, including those involving imperfections, signal variability, and dynamic conditions.

Topological modes (TMs)[1–4]—characterized by their robustness against disorder and backscattering—have facilitated significant advances[5–7] across diverse physical platforms, including photonics[8–10], acoustic[11,12], exciton–polaritons[13,14], cold atoms[15], circuits[16], and mechanical systems[17–19]. These modes are governed by the bulk–boundary correspondence, a fundamental principle stating that the topological invariants of the bulk dictate the existence of localized states at system boundaries. Despite their robustness, the inherent localization of TMs imposes key limitations: it restricts multifunctionality, hinders scalability to complex systems, and limits high-throughput information transport. Achieving spatially designable TMs, without compromising their topological protection, has thus emerged as a central goal in topological physics.

Ongoing efforts have been paid to the interplay between nonlinearity and topological physics, opening the research field of nonlinear topology [20,21]. On the one hand, nonlinearity is ubiquitous in many physical systems. However, the intrinsic locality of nonlinearity disrupts the spatial periodicity of the system, posing challenges for the theoretical understanding of topology [22,23]. On the other hand, nonlinearity and topology combine to give rise to exciting physics and novel phenomena that far exceed the sum of their parts, such as topological solitons [24–27], nonlinearity-induced topological phase transitions [28–34], quantized nonlinear Thouless pumping [35–38], and topological quantum light sources [39–41]. Clearly, nonlinear topology aims to synergize the respective advantages of nonlinearity and topology, creating benefits far greater than the sum of their parts, and paving the way for the next generation of reconfigurable topological devices. Further exploration of nonlinear topological physics remains an intriguing frontier yet to be fully unveiled.

Unfortunately, the nonlinear topological modes (NTMs) underlying these advances are confined to edges or defects and decay into the bulk. As a result, existing NTM-based devices are intrinsically bulky and costly, which hinders scalability to complex architectures and limits their capacity for high-throughput information transport. Recently, arbitrarily reconfigurable nonlinear topological modes (ANTMs) have been proposed [42], these modes are robust against disorders as protected by a nontrivial topology while ,uniquely, can be controllably designed and reshaped into arbitrary profiles through external sources as inherited from the reconfigurability of nonlinearity. These two seemingly mutually exclusive properties greatly enhance topological mode capacity and high-throughput information transport, and are thus highly desirable for nonlinear topological applications. Nonetheless, ANTMs have not yet been demonstrated in any experiment—an essential step in transforming theory into practical applications—as seen in other milestones of topological physics, such as the non-Hermitian skin effect (theory: Ref.[43]; experiment: Ref.[44]), topological laser (theory: Ref.[45]; experiment: Ref.[46]), and non-Hermitian extended modes (theory: Ref.[47]; experiment: Ref.[19]), all of which have significantly advanced the field by supporting theoretical models with vivid experimental demosntrations.

Here, we experimentally demonstrate ANTMs, showing how nonlinearity can be harnessed to deform, reshape, and design the wavefunctions of NTMs. Our circuit integrates a linear topologically nontrivial Su–Schrieffer–Heeger (SSH) chain and a nonlinear counterpart, featuring alternating linear and nonlinear couplings. In the low-power regime, the nonlinear section is topologically trivial, supporting a single localized mode at the interface. As the input power increases, this mode expands into arbitrarily shaped plateaus, such as flat, step-like, "V," and "U" profiles. In our setup, the excitation frequency is fixed, thus avoiding the intensity-dependent tuning typical of previous nonlinear topological studies and greatly simplifying implementation. We further demonstrate real-time transitions between different plateaus. These findings advance the fundamental understanding of nonlinear topology and establish a new paradigm for robust, adaptive wave control. Our results pave the way for next-generation topological devices that are not only topologically protected and resilient, but also programmable and responsive to real-time environmental changes.

**Nonlinear circuits and theoretical analysis**

The basic geometry of our set-up is shown schematically in Fig. 1**a**, and consists of a nonlinear Su–Schrieffer–Heeger (SSH) chain[28] coupled to a linear chain[48]. The nonlinear tight-binding Hamiltonian reads

$$H_{|\psi\rangle} = \sum_{i<n} v_i |a_i\rangle\langle b_i| + \kappa_{i-1}|a_i\rangle\langle b_{i-1}| + \kappa_d \langle a_{n+1}\rangle\langle b_n| + \sum_{i>n} t|a_i\rangle\langle b_i| + \tau|a_{i+1}\rangle\langle b_i| + H.c.. \quad (1)$$

Here $|\psi\rangle \equiv (\cdots, a_i, b_i, \cdots)^T$ is the state with $a_i$ and $b_i$ being the amplitudes on the corresponding sublattices of the $i$-th unit cell. The summation over $i < n$ denotes a nonlinear SSH chain with linear coupling $\kappa_i$ (red line) and nonlinear coupling $v_i$ (blue line). The summation over $i > n$ represents a linear SSH chain with intercell coupling $\tau$ (pink lines) and intracell coupling $t$ (cyan lines) where $\tau > t$. These two chains are coupled through $\kappa_d$ (black line). $H_{|\psi\rangle}$ is realized using the circuit shown in Fig. 1b. The nonlinear coupling between the two LC resonators is realized through a linear capacitor $C_{\tilde{v}}$ and a nonlinear capacitor $C_V$. The voltage dependence of $C_V$ at the operating frequency of 190 kHz is shown in Fig. 1c. The parameters were carefully chosen [see Supplementary Table 1] and $C_V + C_{\tilde{v}} > C_\kappa$ at small wave amplitude and $C_V + C_{\tilde{v}} < C_\kappa$ at large wave amplitudes ($V_{i,A}$ and $V_{i,B}$). [See Methods and Supplementary Information section 1 for the correspondence between the Hamiltonian and the circuit.]

We start with a simple case where the ANTM is designed to be flat in the nonlinear part under a selected intensity. The eigenvalues and eigenstates of the nonlinear Schrödinger equation in Eq. (1) are numerically obtained using a self-consistent method [32]. Figure 1d shows the eigenspectrum versus the total intensity $I = \langle\psi|\psi\rangle$ of the eigenstates, where the red dots mark the ANTMs inside the gap. When $I$ is small, the ANTM is localized at the interface and exponentially decays into both chains (see Fig. 1e at $I = 0.5^2$). As $I$ increases, the wavefunctions inside the nonlinear regime deviate from exponential decay (see Fig. 1e at $I = 8^2$), gradually extend to a plateau with constant amplitude (see Fig. 1e at $I = I_0$), and eventually start to concentrate at the left boundary of the nonlinear lattice (see Fig. 1e at $I = 23^2$).

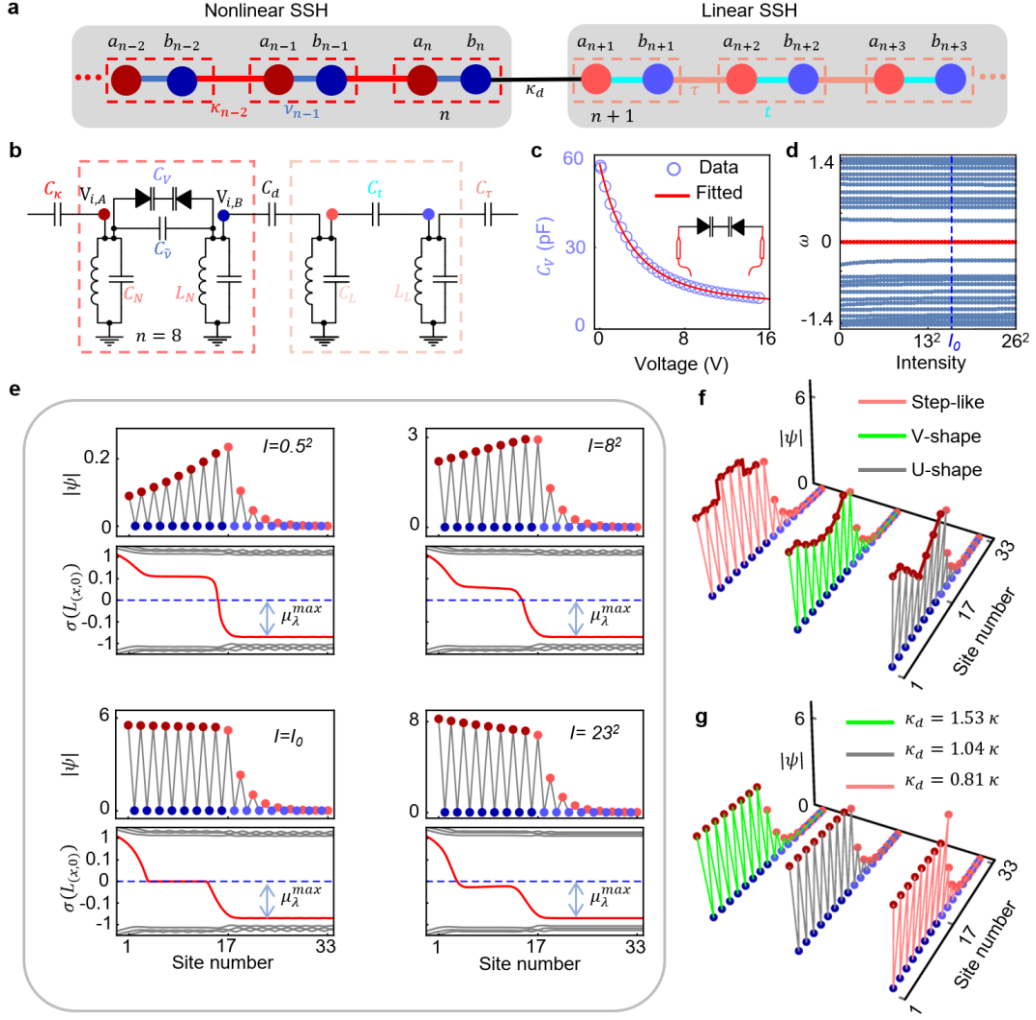

**Fig. 1| ANTMs realized with an electronic circuit. a**, Schematic of a tight-binding model consisting of a nonlinear chain and linear SSH chain with field amplitude $a_i$, $b_i$ of the $i$-th unit cell. **b**, Circuit implementation of the tight-binding model in the experiment. **c**, Measured (open disks) capacitance as a function of voltage applied to the nonlinear capacitor $C_V$. These data points can be approximated by a fitting curve (red line) $C_V(V) = a/(1 + V/b)^c + d$ with fitting parameters: a = 52 pF, b = 8.5 V, c = 2.8 and d=8.5 pF. **d**, Eigenspectra of the effective tight-banding Hamiltonian for a finite lattice composed of 33 sites versus the intensity of the eigenstates, where the red dots mark the ANTMs. **e**, Wavefunctions of the ANTMs and eigenvalues of the reduced spectral localizer for different $I$. In the upper panels, the red and blue dots represent the wavefunction at the $a$ and $b$ sublattices, respectively. In the lower panels, the solid red lines represent the eigenvalues closest to zero. At $I_0 = 16.5^2$, the wavefunction of the ANTM extends to a plateau with a constant amplitude. **f, g,** The wavefunctions of the ANTMs at corresponding $I_0$ for different distribution of $\{\tilde{v}_i\}$ (**f**) and different $\kappa_d$ (**g**). The details of the effective tight-binding Hamiltonian are provided in Supplementary Information section

1. $\beta = 0.5$ in **e**. and the corresponding distribution of $\{\tilde{v}_i\}$ in **f** are provided in Supplementary Information section 3.

The nonlinear dependence of the coupling coefficients breaks translational symmetry, rendering the conventional topological invariant ill-defined in momentum space. Here, we adopt a nonlinear spectral localizer [49,50] to uncover the topological origin of the ANTMs. Since all couplings are real in Eq. (1), its spectral localizer can be written in a reduced form as

$$L_{\lambda \equiv (x,\tilde{\omega})}(X, H_{|\psi\rangle}) = \beta(X - x\mathbf{I})\Pi + H - i\tilde{\omega}\Pi. \qquad (2)$$

Here, $\beta$ is a hyperparameter used to ensure consistency between the units of the position operator $X$ and Hamiltonian $H$, $\mathbf{I}$ is the identity matrix and $\Pi$ is the system's chiral operator. The lines in the lower panel of Fig. 1**e** show the eigenvalues $\sigma(L_\lambda)$ of the reduced spectral localizer at different intensities. When one of $\sigma(L_\lambda)$ crosses zero, such as the red lines, at a location $x$ and frequency $\tilde{\omega}$ [denoted by $\lambda \equiv (x, \tilde{\omega})$], it indicates the existence of a state approximately localized near $\lambda$. If one of $\sigma(L_\lambda)$ is near zero over a finite range of $x$ (lower left panel of Fig. 1**e**), the ANTM becomes extended. The topological protection of the ANTMs guarantees the existence of a ANTM with a similar wavefunction in the presence of randomness (see Methods and Supplementary Information section 2). To show the capability of arbitrarily configuring the TMs, Fig. 1**f** illustrates three different plateau shapes: step-like, U-shaped, and V-shaped. The systematic approach for designing the shape of the plateau is provided in the Supplementary Information section 3. Furthermore, the magnitude of the ANTMs' tails in the linear chain can also be tuned with $\kappa_d$ (see Fig. 1**g**).

**Experimental observation of the ANTMs**

Compared with linear topology, ANTMs depend on the intensity, which offers a unique controllability utilizing external sources. To observe the stable excited ANTMs, we introduce external sources (yellow spots) and losses (unavoidable in nature) as sketched in Fig. 2a. The entire system contains only a single excitation source, located in the first resonator of the linear chain, which significantly simplifies the complexity of the experimental circuit. Figure 2**b** shows the corresponding circuit near the interface. Here, the external source is generated by an arbitrary waveform generator (AWG) with complex voltage $V_e$, working frequency $f = 190$ kHz, and is connected to the circuit through a small resistor $R_e = 50\ \Omega$. There is unavoidable loss (resistors) in real circuits as indicated by the arrows in Fig. 2**a**. Additional loss mechanisms, including parasitic resistances, can be effectively accounted for using equivalent shunt resistors, namely $R_{NA}, R_{NB}, R_{LA}, R_{LB}$ in Fig. 2**b**. See more details in Supplementary Information section 4. Due to the presence of loss, the mode amplitudes on the B sublattices are no longer zero. In principle, the circuit components, particularly the inductors, are subject to variations of the applied voltage and operating frequency. To minimize the voltage dependence of the inductance, we custom-fabricated several inductors, each consisting of 15 meters of bare copper wire (composed of 47 strands, each with a 1.5 mm² cross-sectional area) wound around a 10 cm-diameter PVC cylinder, as shown in Fig. 2**c**. A more detailed experimental setup is provided in Supplementary Information section 5. Figure 2**d** shows the dynamics of $V_{i,A}$ in the nonlinear circuit for stable excited states under two different $|V_e|$. Notably, these $V_{i,A}$ cross zero simultaneously, an exclusive feature of the ANTMs in our set-up. The amplitude of voltage on each sublattice, $|V_{i,A}|$ and $|V_{i,B}|$, is extracted from the measured dynamics. We define the intensity of the excited ANTM as $I_e = \sum_i \left( |V_{i,A}|^2 + |V_{i,B}|^2 \right)$. Figure 2**e** shows $I_e$ versus $|V_e|$, where the open disks represent the measured data and the red lines are derived from the coupled mode theory (CMT, details in Supplementary

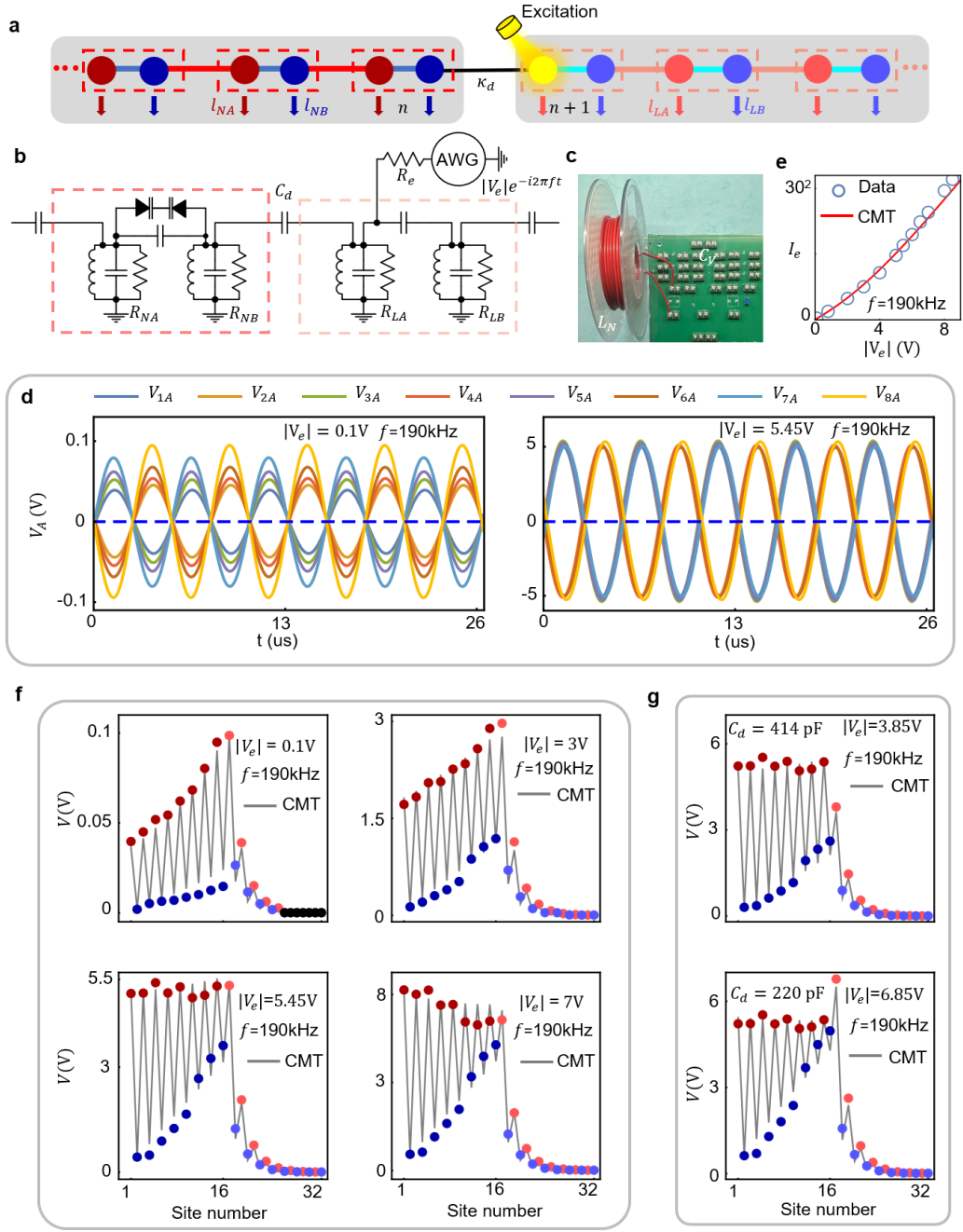

**Fig. 2 | Observation of ANTMs in a circuit. a**, Schematic of excitation. The source only acts on the first site (yellow spot) of the linear part, and the arrows indicate the intrinsic loss. **b**, Circuit implementation of the model in **a**. The arbitrary waveform generator (AWG) is used to excite the ANTMs through a resistor $R_e = 50\ \Omega$. **c**, A photo of a home-made inductor $L_N$ and part of the printed circuit board (PCB). **d**, Measured temporal dynamics of $\{V_{iA}\}$ for two different $|V_e|$. **e**, The total intensity of the excited stable state $I_e$ versus the output voltage of the AWG, $|V_e|$. **f**, **g**, Measured amplitudes of the ANTMs for different $|V_e|$ and $|C_d|$. The color coding of the dots follows the same convention as in Fig. 1. The lines in **d**, **f**, **g** are obtained with the coupled mode theory (CMT). In the

tight-binding calculations, the parameters used are consistent with those in Fig. 1. The black dots in the upper-left panel of **f** indicate that the measured voltage falls below the experimental noise floor (approximately 3 mV) and is therefore set to zero. In the experiments, the working frequency is fixed at $f = 190\text{kHz}$. Unlike previous studies on nonlinear topology [29]—where the operating frequency required continuous tuning with the driving voltage because both the cavity resonance and coupling strength depend on intensity [30,31,51]—our fixed excitation frequency avoids this complication. This makes practical implementations far simpler, highlighting the desirability of a fixed frequency in nonlinear topological systems. Details of the circuit elements on the PCB and the corresponding CMT are provided in Supplementary Information section 4 and section 5.

Information section 4). It is clear that there is a one-to-one correspondence between $I_e$ and $|V_e|$, which is crucial for using $|V_e|$ as an external knob. Figure 2**f** shows the specific waveform of the excited ANTMs under different $|V_e|$. The corresponding waveform also undergoes a deviation ($|V_e| = 3V$) from exponential decay ($|V_e| = 0.1V$), gradually extends to a plateaus with near-constant-amplitude ($|V_e| = 5.45V$), and eventually concentrates at the nonlinear boundary ($|V_e| = 7V$). Figure 2**g** shows the measured voltages when the excited ANTMs are nearly flat in the nonlinear circuit part for different $C_d$, which shows that one can adjust $C_d$ to tune the wave amplitudes in the linear circuit. Despite inevitable variations in component values, the excited ANTMs remain stable, highlighting their robustness against parameter imperfections. For instance, the capacitors used in our circuits were not preselected and exhibit a standard variation of 5%. Additionally, no extra fine-tuning was applied to any circuit elements on the PCB. Despite such a high level of error, the experimental demonstration in Fig. 2 remains reasonably good. More detailed error analysis is presented in Supplementary Information section 6.

**ANTMs exhibiting other plateaus**

By varying the parameters of the system, the ANTMs can be designed to exhibit arbitrary shapes. In the CMT, the ANTMs are governed by the following equation:

$$\begin{aligned}
(\tilde{v}_i + f(a_i, b_i))a_i + \kappa_i a_{i+1} &= 0, \quad (i < n) \\
(\tilde{v}_i + f(a_i, b_i))a_i + \kappa_d a_{i+1} &= 0, \quad (i = n) \\
t a_i + a_{i+1}\tau &= 0. \quad (i > n)
\end{aligned} \tag{3}$$

Here $f(a_i, b_i)$ represents the nonlinear response of the coupling coefficient that depends on $a_i$ and $b_i$ (see Methods). Clearly, the plateau of the ANTMs (i.e., the distribution of $\{a_i\}$ at the corresponding intensity $I_0$) is determined by $\{\tilde{v}_i\}$, $\{f(a_i, b_i)\}$ and $\{\kappa_i\}$. Given the shape of a designed plateau, one can solve for the corresponding distribution of $\{\tilde{v}_i\}$ with Eq. (3). In the circuit, $\{\tilde{v}_i\}$ corresponds to the $\{C_{\tilde{v}i}\}$. For demonstration purposes, we design three common shapes: a step-like, a V-shape, and a U-shape. Implementing the same excitation (190 kHz) as before, fig 3**a** illustrates the evolution of the excited stable state as a function of $|V_e|$ for different designs. At low-power excitation (first column), these wavefunctions of ANTMs are localized at the interface and decay into both chains. As $|V_e|$ increses, the wavefunctions deviate from the initial decay, subtly revealing traces of the designed plateau (second column) and gradually extending to the intended plateau (third column), eventually tending to concentrate at the left boundary (fourth column). These experimental results are in excellent agreement with the theoretical analysis.

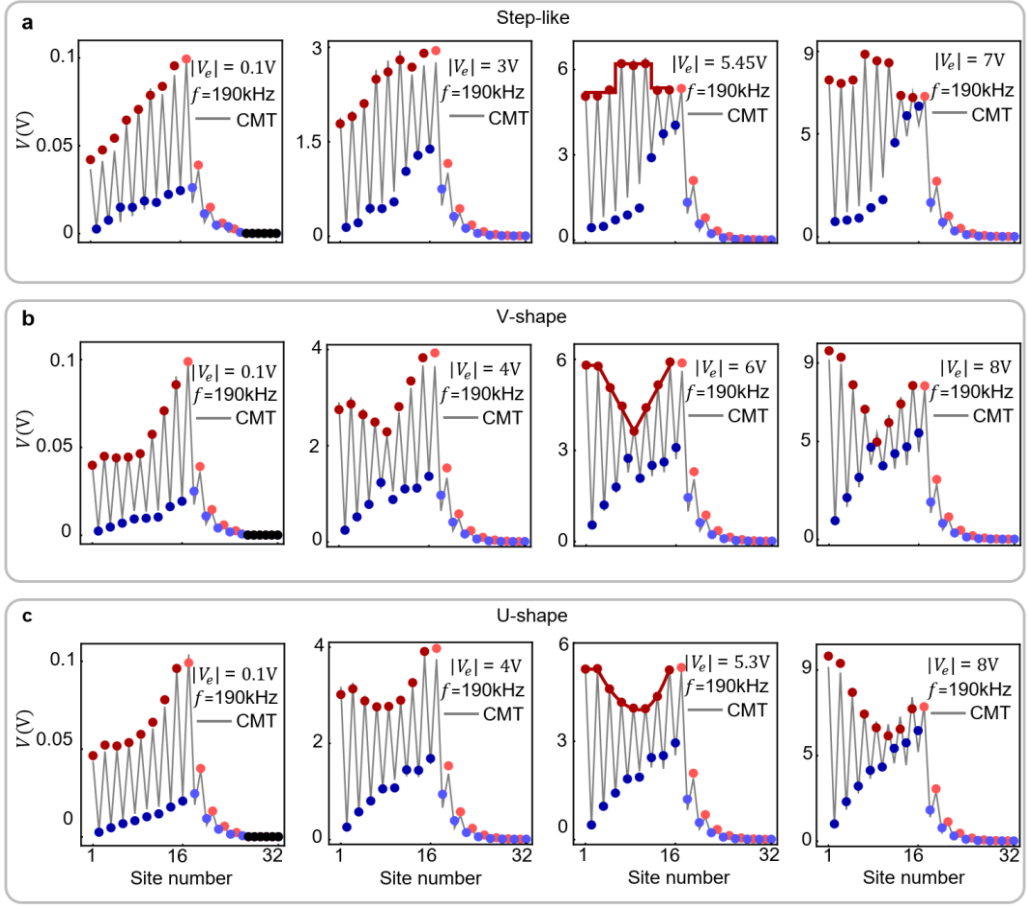

**Fig. 3 | Design the plateaus of the ANTMs.** Similar to Fig. 2**f**, but with redesigned $\{C_{\tilde{v}i}\}$ to demonstrate the reconfigurability of the ANTMs. The plateaus are designed as a step-like profile in **a**; a V-shape in **b**; and a U-shape in **c**. At a certain $V_e$ (see the third column), these wavefunctions exhibit the designed plateau, with the contours outlined with the solid red lines. The black dots represent voltages below the experimental noise floor. The corresponding capacitance distributions $\{C_{\tilde{v}i}\}$ are provided in Supplementary Fig.9, and the other unspecified parameters are the same as Fig. 2.

**Real-time transitions between different plateaus of ANTMs.**

Given the predefined parameters $\{\tilde{v}_i\}$, the shape of the plateau (third column) is determined. The real-time transitions of the ANTMs' plateaus invigorates research in nonlinear topology and unlocks new possibilities for potential applications. To further clarify the theoretical mechanism underlying these real-time transitions, we take photonic systems as an illustrative example. We can introduce the pumping beam and signal beam as have already been implemented in Refs [34,51,52]. Generally, the nonlinear components exhibit a response time that is much longer than the phase variation time scale of the pumping beam and signal beam[53]. Thus, the time-averaged intensities at different sublattices of the *i*-th unit cell can be given as $I_{a_i} = |a_{i,s}|^2 + |a_{i,p}|^2$ and $I_{b_i} = |b_{i,s}|^2 + |b_{i,p}|^2$ [see Eq.(1) in Ref [53]]. Now, the nonlinear coupling can be extended as

$$v_i = \tilde{v}_{i0} + f(a_{i,p}, b_{i,p}) + f(a_{i,s}, b_{i,s}). \tag{4}$$

Consequently, the reshaping of signal plateaus can be reconfigured by real-time tuning the profile of the pump beam. However, several challenges still remain. Firstly, to control the profile of the ANTMs in real-time, it is essential to solve for all possible ANTMs under the parameters variations (including different excitation source intensity and the distribution of $\{\tilde{v}_i\}$), thereby confirming that the system does not reach other unintended stable states. Secondly, we need to ensure the stability of the ANTM during its dynamical evolution. Thirdly, it is necessary to estimate the response time required for the system to reach the target ANTM. After addressing these challenges, we enable shape-shifting between arbitrarily designed plateaus (see Supplementary Information, Section 7, for a detailed analysis). To provide a more intuitive demonstration of the real-time transitions between different plateaus, we introduce a voltage-controlled tunable capacitor $C_{\tilde{v}1}$ to serve as the pump beam. Consequently, $C_{\tilde{v}}$ consists of a standard capacitor $C_{\tilde{v}0}$ in parallel with a voltage-controlled tunable capacitor $C_{\tilde{v}1}$. Figure 4**a** shows the schematic of the circuit diagram employed to achieve this functionality. The upper part of Fig. 4**b** presents a photograph of the capacitor $C_{\tilde{v}1}$, while the lower panel shows the measurement circuit. By adjusting the applied voltage $V_{DC}$, the tunable capacitance can be precisely modulated, as demonstrated in Fig. 4**c**, which depicts the measured dependence of $C_{\tilde{v}1}$ on voltage $V_{DC}$. We integrate the circuit shown in Fig. 4**a** into the fourth unit cell of the system (i.e., adding an additional capacitor $C_{\tilde{v}1}$) while maintaining the other settings identical to those in Fig. 2. Initially, $V_{DC}$ is set to 3 V, allowing the system to reach a stable state. The corresponding ANTM wavefunction is displayed in Fig. 4**d**, exhibiting a flat plateau (highlighted by the purple solid line). To investigate the transition between different plateau configurations, $V_{dc}$ is switched from 3 V to 0 V at $t = 2$ ms. The resulting evolution of $\{V_{i,A}\}$ is depicted in Fig. 4**e**, where a distinct transformation occurs immediately following the voltage $V_{DC}$ change. By $t = 4$ ms, the system stabilizes into another designed steady state, characterized by a staircase-like plateau structure. The final wavefunction is presented in Fig. 4**f**. It is important to note that, prior to the current transitions, the excitation voltages $|V_e|$ required to reach the plateaus differ only slightly; therefore, we did not perform additional tuning, in contrast to

Fig. 2g. This demonstration highlights the feasibility of dynamically modulating ANTMs, paving the way for real-time adaptability in nonlinear topological systems.

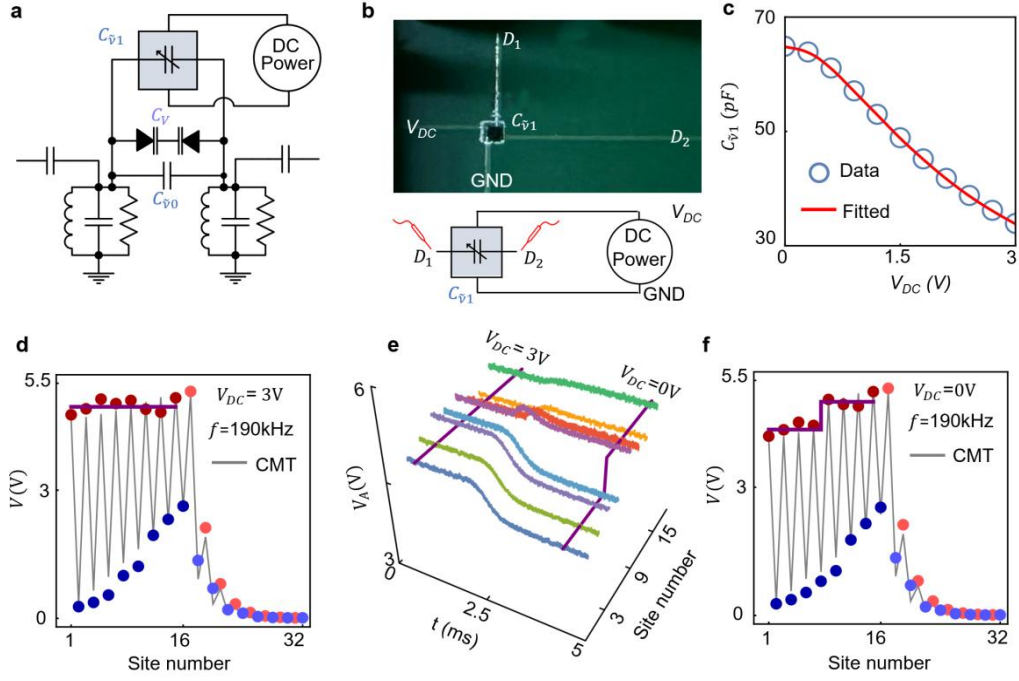

**Fig. 4| Real-time transitions between different plateaus of ANTMs. a**, Circuit implementation used in our experiment to achieve real-time reconfigurability through an external DC power supply. Here, $C_{\tilde{v}}$ defined in Fig. 1c is formed by the parallel combination of a standard capacitor $C_{\tilde{v}0}$ and a voltage-controlled tunable capacitor $C_{\tilde{v}1}$. **b**, A photo of $C_{\tilde{v}1}$ on the PCB along with the circuit used to calibrate its properties (lower panel). **c**, Measured capacitance–voltage relation for $C_{\tilde{v}1}$ (blue circles), with the red fitting curve included as a visual guide. **d**, Measured amplitude of the ANTM at $V_{DC} = 3\,V$, where the circuit in **a** is integrated into the fourth unit cell and the remaining unit cells are identical to those in Fig. 2. **e**, Measured temporal evolution of $V_A$, where $V_{DC}$ is switched from 3V to 0V at around 2 ms. **f**, Measured amplitude of the ANTM at $V_{DC} = 0\,V$, showing the reconfigured state. In **d-f**, the solid purple lines outline the designed plateau, clearly illustrating that real-time reconfiguration of the ANTMs is achieved through tuning $V_{DC}$.

**Summary and Discussions**

In summary, we have experimentally realized ANTMs in electrical circuits, establishing a new class of topological phases that arises uniquely from the interplay between nonlinearity and topology. These modes can be dynamically reshaped in real time by tuning the excitation amplitude, without compromising topological protection—a capability that fundamentally reformulates the conventional understanding of BBC in nonlinear systems. Our platform enables real-time control of spatial mode profiles with arbitrary design profiles, as demonstrated through multiple representative shapes. While implemented in a one-dimensional circuit, the approach is readily extendable to higher-dimensional systems, higher-order topological modes, and scalable platforms such as PCBs and on-chip technologies.

Beyond the demonstration of tunability and robustness, our results open new directions for exploring nonlinear topological phenomena. As detailed in Supplementary Information section 4, ANTMs exhibit exceptional stability under variations in excitation frequency. Our findings also raise compelling questions: Can ANTMs propagate through a lattice akin to solitons[54]? How might periodic driving[55] or non-Hermiticity be harnessed to enrich their dynamics? Could ANTMs enable multistable topological states, enhancing the information capacity of nonlinear devices and enabling new schemes for classical or quantum information processing[56]? We anticipate that the concept of ANTMs will serve as a foundation for programmable, robust topological functionalities across platforms including photonics, acoustics, plasmonics, polaritonics, and ultracold atomic systems.

## Methods

**Realization of nonlinear saturable hopping.**

The implemented circuit consists of the dimer-type unit cells shown in Fig. 1c. In the nonlinear part, the two identical LC resonators within a dimer are coupled through a linear capacitor $C_{\tilde{v}}$ and a nonlinear capacitive composite $C_V$. $C_V$ is realized using two varactor diodes connected in series in a back-to-back configuration. This design eliminates the need for a DC bias, typically required to prevent forward conduction in the diodes. During each half-cycle, one diode is forward biased, effectively acting as a short circuit, while the other is reverse biased, functioning as the nonlinear capacitor. This approach enables the full utilization of the nonlinear dynamic range of the diodes, thereby maximizing the nonlinear effect. Here, we employ a back-to-back design using SkyWorks SMV1470 diodes. The inset of Fig. 1b illustrates the measurement setup. Under a 190 kHz cosine test signal, the capacitance $C_V$ decreases from 60.5 pF to 8.5 pF as the test voltage is varied from 0 to 16 V. Thus, when the implemented circuit operates a steady-state regime, with an operating frequency near 190 kHz, the capacitance $C_V$ in the $i$-th cell can be given by

$$C_V = a / \left(1 + \sqrt{|V_{i,A}|^2 + |V_{i,B}|^2 - 2|V_{i,A} V_{i,B}| \cos(\phi_i)}/b\right)^c + d, \tag{4}$$

where $\phi_i$ represents the phase difference between the complex voltages $V_{i,A}$ and $V_{i,B}$. The other parameters used are $a$ = 52 pF, $b$ = 8.5 V, $c$ = 2.8 and $d$=8.5 pF. Using the CMT, and normalizing all parameters by the intercell coupling $\kappa$ (implemented by $C_\kappa$), the intracell coupling in the nonlinear cell can be mapped to a saturable nonlinear coupling as

$$v_i = \frac{C_{\tilde{v}} + d + a}{C_\kappa} + \frac{a/\left(1 + \sqrt{|a_i|^2 + |b_i|^2 - 2|a_i b_i| \cos(\theta_i)}/b\right)^c - a}{C_\kappa} \tag{5}$$
$$\equiv \tilde{v}_i + f(a_i, b_i).$$

**Topological protection of the ANTMs.**

An intuitive physical insight into the connection between the spectral localizer[49,50] and the topology of the material can be drawn from its behavior at the atomic limit. A material is topologically nontrivial if it cannot be continuously deformed into an atomic limit without either closing a gap or breaking a symmetry. In the atomic limit, $X$ and $H$ commute. In this case, the signature of the spectral localizer sig($L_\lambda$), is zero. Here, the signature is defined as the number of eigenvalues with positive real parts minus those with negative real parts. If sig($L_\lambda$) $\neq$ 0, there exists an obstruction to deforming the system into the atomic limit where $X$ and $H$ commute. This implies that the system exhibits a topologically nontrivial phase at $\lambda$. When sig($L_\lambda$) changes, one eigenvalue of $\sigma(L_\lambda)$ crosses 0 (as illustrated by the red lines in the lower panel of Fig. 1**e**). This indicates the existence of a state approximately localized near $\lambda$, thus realizing the unique BBC in our nonlinear lattice. The smallest singular value $\mu_\lambda$=min[|$\sigma(L_\lambda)$|] of the spectral localizer can also provide additional information about the system at $\lambda$. Values of $\mu_\lambda$ close to zero indicate the presence of a state approximately localized near $\lambda$, whereas large ones indicate that the system does not support such a state. Consequently, $\mu_\lambda$ can be interpreted as a "local band gap", and the topological protection of the ANTMs can be characterized by

$$||\Delta H(\delta)|| \leq \mu_\lambda^{max}, \qquad (6)$$

where $||\Delta H(\delta)||$ is the largest singular value of $\Delta H(\delta) \equiv H_{|\psi\rangle}(\delta) - H_{|\psi\rangle}$, with $H_{|\psi\rangle}(\delta)$ representing the perturbation on the nonlinear Hamiltonian; $\mu_\lambda^{max} \equiv max_x[\mu_{(x,0)}]$ denotes the maximum $\mu_{(x,0)}$ inside the topological domain. As long as Eq. (6) is satisfied, the topological protection guarantees the existence of a ANTM with a similar wave function.

**Measurement instruments.**

The inductances, capacitances, and resistances were experimentally measured using precision LCR meters (TH2829C and Keysight E4980A). Excitation signals were applied using a Keysight 33600A arbitrary waveform generator, and the responses were recorded with a Keysight MXR608B oscilloscope.

**Data availability**

The data that support the plots within this paper and other findings of this study are available from the corresponding author upon reasonable request.

**Acknowledgments**

This work is supported by the National Key Research and Development Program of China (Grant No. 2022YFA1404900), the National Natural Science Foundation of China (Grant No. 12334015, 12274330, 12274332, 12321161645,12404440), the Key Research and Development Program of the Ministry of Science and Technology (Grants No. 2024YFB2808200), the China Postdoctoral Science Foundation under Grant No. 2023M742715, the China National Postdoctoral Program for Innovative Talents under Grant No. BX20240266 and Postdoctor Project of Hubei Province under Grant No. 2024HBBHCXB054.

**Author contributions**

M.X., D. W. and K.B. initiated the project. K.B. did the simulations, designed the samples and performed the experiments with help from C.L., J.-Z.L, and T.L. All authors contributed to discussions of the results. M.X., and K.B. wrote the manuscript.

**Competing interests**

The authors declare no competing interests.